\documentclass[sigconf]{acmart}

\usepackage{tcolorbox}
\tcbuselibrary{theorems}

\usepackage{enumerate}
\usepackage{graphicx}
\usepackage{algorithmic}
\usepackage{subcaption}
\usepackage{amsmath,amsfonts,amsthm}
\usepackage{algorithmic}
\usepackage{graphicx}
\usepackage{textcomp}
\usepackage{soul}
\usepackage{colortbl}
\usepackage{tikz}
\usepackage{enumitem}
\usetikzlibrary{arrows.meta,
                chains,
                positioning,
                shapes.geometric,
                shapes.symbols,
                decorations.pathreplacing,
                calligraphy
                }
\usepackage{comment}
\usepackage{fancyhdr}
\usepackage[english]{babel}
\usepackage{multirow}
\usepackage{multicol, blindtext}
\usepackage{diagbox}
\usepackage{hyperref}
\usepackage{fancyhdr}
\usepackage[ruled,linesnumbered,vlined,norelsize]{algorithm2e} 

\SetCommentSty{mycommfont}
\SetAlFnt{\footnotesize\sf}
\newcommand{\RN}[1]{%
  \textup{\uppercase\expandafter{\romannumeral#1}}%
}

\usepackage{enumitem}
\usepackage{forest}
\usetikzlibrary{shadows}
\theoremstyle{definition}

\setstcolor{red}
\newcommand{\red}[1]{\textcolor{red}{#1}}
\newcommand{\blue}[1]{\textcolor{blue}{#1}}

\newcommand{\caroline}[1]{\textit{\color{blue}[Caroline: #1]}}

\usetikzlibrary{shapes}
\usepackage{multirow}
\usepackage{booktabs}
\usepackage{tikz} 

\usetikzlibrary{arrows.meta,
                chains,
                positioning,
                shapes.geometric,
                shapes.symbols
                }

\newcommand*\emptycirc[1][1ex]{\tikz\draw (0,0) circle (#1);} 
\newcommand*\halfcirc[1][1ex]{%
  \begin{tikzpicture}
  \draw[fill] (0,0)-- (90:#1) arc (90:270:#1) -- cycle ;
  \draw (0,0) circle (#1);
  \end{tikzpicture}}
\newcommand*\fullcirc[1][1ex]{\tikz\fill (0,0) circle (#1);} 

\usepackage[normalem]{ulem}
\usepackage{graphicx}

\usepackage{rotating}
\newcommand{\rotcell}[1]{
  \rotatebox[origin=c]{45}{ #1 }}
\newcommand{\myrotcell}[1]{\rotcell{\makebox[0pt][l]{\small{#1}}}}

\theoremstyle{definition}

\definecolor{mymauve}{rgb}{0.1,0.2,0.7}
\definecolor{mypurple}{rgb}{0.8, 0.2, 0.6}
\definecolor{olivebrown}{cmyk}{.6,.4,0.8,0}
\definecolor{olivegreen}{rgb}{0.0, 0.5, 0.0}

\newtcbtheorem{mytheo}{Definition}%
{colbacktitle=violet!75,colframe=black}{th}

\newlength\Origarrayrulewidth

\AtBeginDocument{%
  \providecommand\BibTeX{{%
    \normalfont B\kern-0.5em{\scshape i\kern-0.25em b}\kern-0.8em\TeX}}}

\setcopyright{acmcopyright}
\copyrightyear{2024}
\acmYear{2024}
\acmDOI{10.1145/1122445.1122456}

\acmConference[SE 2030]{International Workshop on Software Engineering in 2030}{November 2024}{Puerto Galinàs (Brazil)}
\acmPrice{15.00}
\acmISBN{978-1-4503-XXXX-X/18/06}

\author{Sinclair Hudson}
\email{sinclair@cs.toronto.edu}
\affiliation{%
  \institution{University of Toronto}
  \city{Toronto}
  \state{Ontario}
  \country{Canada}
}

\author{Sophia Jit}
\email{sophia.jit@mail.utoronto.ca}
\affiliation{%
  \institution{University of Toronto}
  \city{Toronto}
  \state{Ontario}
  \country{Canada}
}

\author{Boyue Caroline Hu}
\email{boyue@cs.toronto.edu}
\affiliation{%
  \institution{University of Toronto}
  \city{Toronto}
  \state{Ontario}
  \country{Canada}
}

\author{Marsha Chechik}
\email{chechik@cs.toronto.edu}
\affiliation{%
  \institution{University of Toronto}
  \city{Toronto}
  \state{Ontario}
  \country{Canada}
}

\renewcommand{\shortauthors}{Sinclair Hudson, Sophia Jit, Boyue Caroline Hu, and Marsha Chechik}

\begin{abstract}

Large Language Models (LLMs) are rapidly becoming ubiquitous both as stand-alone tools and as components of current and future software systems.  To enable usage of LLMs in the high-stake or safety-critical systems of 2030, they need to undergo rigorous testing. 
Software Engineering (SE) research on testing Machine Learning (ML) components and ML-based systems has systematically explored many topics such as test input generation and robustness.  We believe knowledge about tools, benchmarks, research and practitioner views related to LLM testing needs to be similarly organized.  To this end, we present a taxonomy of LLM testing topics and conduct preliminary studies of state of the art and practice approaches to research, open-source tools and benchmarks for LLM testing, mapping results onto this taxonomy.
Our goal is to identify gaps requiring more research and engineering effort and inspire a clearer communication between LLM practitioners and the SE research community.
\end{abstract}

\title{A Software Engineering Perspective on Testing Large Language Models: Research, Practice, Tools and Benchmarks}

\begin{document}


%

\keywords{LLMs, Software Engineering for Machine Learning, Testing}

\maketitle

\section{Introduction} %
\label{sec:introduction}



Large Language Models (LLMs) are a type of Machine Learning (ML) designed to understand and generate human-like text. They are gaining interest from both researchers and practitioners due to their outstanding performance in various domains, such as language translation and medical text analysis~\cite{chang2023survey}.  Given the growing usage of LLMs as components in software systems, the question of how to specify the expected properties of such systems (e.g., correctness or fairness) and effectively test them becomes very important.  Given the speed with which LLMs have taken center stage, successes and gaps in LLM testing approaches and practices have not yet been systematized in a clear, principled, and comprehensive way, leading to difficulties in applying them in practice~\cite{challengesllmdeployment}.  

LLMs are a particular type of ML model, and testing of ML models has been studied extensively in software engineering literature.  Specifically, Zhang et. al~\cite{ZhangHML22} proposed a taxonomy of ML testing, with the following categories:  (1) testing workflows (how to test), e.g., test oracle generation; (2) testing components (where to test), e.g., data and learning program testing; (3) testing properties (what to test), e.g., privacy and fairness; and (4) application scenarios, e.g., autonomous driving (see  Fig.~\ref{fig:taxonomy}). 

%



\renewcommand\fbox{\fcolorbox{black}{gray!10}}

\begin{center}
    \scalebox{0.95}{\fbox{%
    \parbox{\linewidth}{%
        \textbf{Our vision.} The current research and industrial practices on LLM testing should be organized through the SE lens, i.e., organizing existing methods with topics of ML testing in SE. This organization should facilitate the identification of gaps and support communication among stakeholders, including LLM testing researchers, tool developers, and users. Ultimately, this should contribute to the safer usage of LLMs in software in high-stake domains by 2030.
    }%
}
}
\end{center}


To validate our vision, we present a preliminary study of current research methods, benchmarks, testing tools, and industry practices for LLM testing, aiming to answer two research questions: (\textbf{RQ1}:) To what extent do current open-source benchmarks, tools, and online discussions addressing LLM testing cover research topics in the ML testing taxonomy? (\textbf{RQ2}:) What are the remaining gaps preventing practitioners from effectively applying LLM testing research? 

The rest of this paper is organized as follows: We first extend the ML testing taxonomy of \cite{ZhangHML22} with LLM-specific research topics and then identify topics in ML testing not yet addressed by LLM testing (Sec.~\ref{sec:llm_research}).  Using this taxonomy, we then provide a brief survey of open-source benchmarks and tools (Sec.~\ref{sec:benchmarks} and \ref{sec:tools}) and
 present a preliminary examination of LLM testing in practice through discussions on  Reddit, where practitioners share their experiences with testing LLMs as well as post and answer questions about testing (Sec.~\ref{sec:developers}). We summarize our findings and our vision in Sec.~\ref{sec:future}.
\vspace{-0.1in}
\section{LLM Testing Research}
\label{sec:llm_research}

    \tikzset{
        my node/.style={
            draw=gray,
            inner color=white,
            outer color=white,
            thick,
            minimum width=1cm,
            text height=1.5ex,
            text depth=0ex,
            font=\sffamily,
            drop shadow,
        }
    }

    \begin{figure}[t]
        \centering
\scalebox{0.55}{
    \begin{forest}
        for tree={%
            my node,
            l sep+=5pt,
            grow'=east,
            edge={gray, thick},
            parent anchor=east,
            child anchor=west,
            edge path={
                \noexpand\path [draw, \forestoption{edge}] (!u.parent anchor) -- +(10pt,0) |- (.child anchor)\forestoption{edge label};
            },
            if={isodd(n_children())}{
                for children={
                    if={equal(n,(n_children("!u")+1)/2)}{calign with current}{}
                }
            }{}
        }
        [ML Testing
        [Testing workflow (how to test)
        [Benchmarks,  draw=blue
        [Composite benchmarks, color=blue]
        [Multi-modal benchmarks, color=blue]
        [Specific benchmarks, color=blue]
        [General benchmarks, color=blue]
        ]
        [Debug and repair]
        [Bug report analysis]
        [Test prioritization and reduction]
        [Test adequacy generation]
        [Test oracle identification, outer color=green!20, draw=blue]
        [Test input generation, draw=blue]]
        [Testing components (where to test)
        [Framework testing]
        [Learning program testing, draw=blue]
        [Data testing]]
        [Testing properties (what to test)
        [Correctness, outer color=green!20, inner color=orange!70, draw=blue [Test for Hallucination,  outer color=green!20]]
        [Privacy]
        [Interpretability]
        [Fairness, draw=blue]
        [Efficiency]
        [Robustness and security, draw=blue]
        [Model relevance, outer color=orange!20]]
        [Application Scenario
        [Natural language inference, draw=blue]
        [Machine translation, draw=blue]
        [Autonomous driving, outer color=gray!20]
        [Natural science \& engineering, color=blue]
        [Social science, color=blue]
        [Medical applications, color=blue]
        [Agent applications, color=blue]
        [Other applications, color=blue]
        ]
        ]
    \end{forest}   }
    
    \textbf{Legend}: \small{ML testing topics, \blue{LLM-specific topics}, }
    \scalebox{0.65}{
    \begin{forest}
        for tree={%
            my node,
            l sep+=5pt,
            grow'=east,
            edge={gray, thick}
        }
        [Covered by LLM research, draw=blue]
    \end{forest}}
    \scalebox{0.65}{
    \begin{forest}
        for tree={%
            my node,
            l sep+=5pt,
            grow'=east,
            edge={gray, thick}
        }
        [Not relevant for LLMs, outer color=gray!20]
    \end{forest}
    \begin{forest}
        for tree={%
            my node,
            l sep+=5pt,
            grow'=east,
            edge={gray, thick}
        }
        [benchmark strengths, outer color=orange!20]
    \end{forest}
    \begin{forest}
        for tree={%
            my node,
            l sep+=5pt,
            grow'=east,
            edge={gray, thick}
        }
        [open-source tool strengths, outer color=green!20]
    \end{forest}}
    \vspace{-0.1in}
    \caption{
Taxonomy of ML testing in SE~\cite{ZhangHML22} with additional LLM-specific topics.
}
    \label{fig:taxonomy}
    \vspace{-0.3in}
    \end{figure}
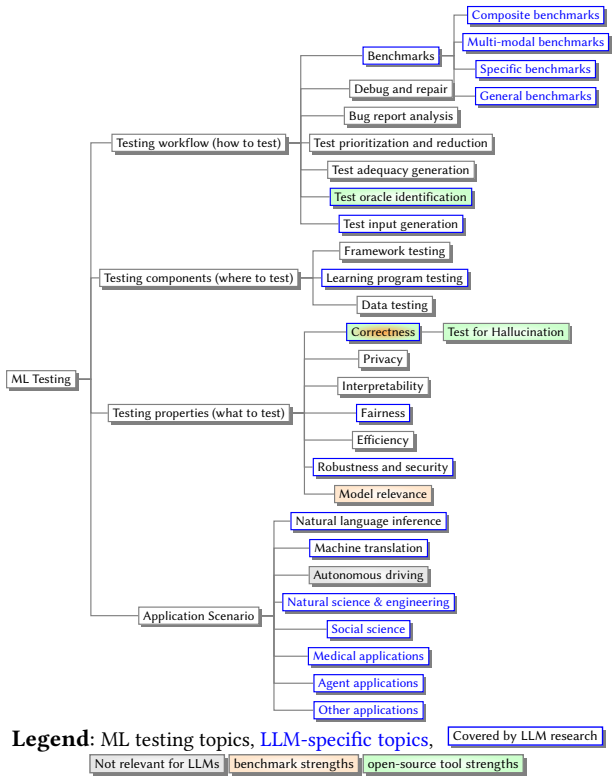
In this section, we extend our taxonomy with research on LLM testing in Fig.~\ref{fig:taxonomy}. Due to the recent drastic increase in LLM-based research, we selected two most recent comprehensive literature reviews on LLM evaluation~\cite{chang2023survey,zhao2023survey}. 

\vspace{0.05in}
\noindent
\textbf{ML Testing Topics Covered by LLM research.} Several  ML testing topics have been explored by LLM research, shown as blue outline in Fig.~\ref{fig:taxonomy}.
For testing workflow, LLM-based applications are commonly evaluated with benchmarks~\cite{chang2023survey}, synthetic data, and human annotations~\cite{zhao2023survey} corresponding to test input and oracle generation. 
Unlike ML testing, existing work on LLM evaluation has primarily focused on learning program testing, i.e., testing trained LLMs, and mostly testing w.r.t. their specific applications~\cite{chang2023survey}, e.g., natural language processing, reasoning, medical usage, etc. Testing properties studied by LLM research include robustness, fairness, and trustworthiness~\cite{chang2023survey}.
It has been shown that LLMs can fabricate false information that sounds plausible, i.e., "hallucinate"~\cite{hallucinationsurvey}. 
As a result, correctness testing also checks whether the model outputs strictly factual information. Several techniques have been developed for this, including factual consistency tests~\cite{summedits, halueval}, uncertainty-based tests~\cite{hallucinationsurvey}, and black-box tests~\cite{selfcheckgpt, adlakha2023evaluating, chatgptfactualeval}.
Hallucination is also a studied failure mode in machine translation, "where translations are grammatically correct but unrelated to the source sentence"~\cite{muller-etal-2020-domain, raunak-etal-2021-curious}. This definition differs slightly from hallucination in the context of LLMs, where there is no source sentence and the truthfulness of the generated text is of principle concern~\cite{hallucinationsurvey}.

\vspace{0.05in}
\noindent
\textbf{New Entries: LLM-specific Topics.} In addition to ``general'' ML testing topics, research in LLM testing has also explored topics specific to LLMs, shown as blue text in Fig.~\ref{fig:taxonomy}. 
First, due to the capabilities of LLMs being suitable for a large variety of tasks, LLMs have been tested for correctness, robustness, fairness, trustworthiness~\cite{chang2023survey}. In addition, LLMs have been deployed in systems for various other domains including medical and social science, or other specific applications such as personality tests~\cite{bodroza2023personality}.



\vspace{0.05in}
\noindent
\textbf{Conclusion.} 
Addressing \textbf{RQ1}, our initial review indicates that several SE testing topics are relevant but have yet to receive sufficient attention in LLM testing research. 

\vspace{0.05in}
\noindent
\textbf{Threats to Validity.} We looked at only two literature reviews, although they were comprehensive and recent at the time of writing.  


\section{LLM Evaluation Benchmarks}
\label{sec:benchmarks}
Research in LLMs has primarily relied on benchmarks for evaluation. Popular LLMs such as GPT-4 and Gemini use multiple benchmarks to communicate and compare their performance across a wide range of tasks and domains~\cite{gpt4, gemini}.  In this section, we analyze the OpenCompass GitHub repository~\cite{opencompass}, a popular collection of 76 publicly available benchmarks, aiming
to identify under-explored LLM testing topics.   OpenCompass was picked as the highest-starred repository of LLM benchmarks found by 
 searching "llm evaluation" in GitHub.

\vspace{0.05in}
\noindent
\textbf{Benchmark Coverage.}
The studied benchmarks included \textit{general benchmarks}, e.g., SQuAD2.0~\cite{squad2} developed for general language understanding, \emph{specific} ones for particular tasks, e.g., CMB~\cite{wang2024cmb} developed for medical applications, and \emph{multi-modal} benchmarks for tasks concerning both images and text, e.g., MMBench~\cite{mmbench} for vision language models.
74 out of 76 benchmarks are dedicated to testing model correctness across different capabilities including mathematical reasoning, content summary, and code generation~\cite{opencompass}.
The remaining two benchmarks, CrowS-Pairs~\cite{nangia2020crows} and AdvGLUE~\cite{wang2022adversarial}, test model fairness and robustness, respectively.
Certain benchmarks such as SciBench~\cite{wang2024scibench}, HumanEval~\cite{humaneval}, and CMB~\cite{wang2024cmb} can be used to assess model relevance to certain application scenarios.
There were no benchmarks in the collection dedicated to privacy, security, efficiency, or interpretability.
Some applications, e.g., open-ended chatbots, require that the LLM be tested on a wide variety of tasks, to see how well it responds.  
\textit{Composite benchmarks} in OpenCompass, such as MMBench~\cite{mmbench} and SummEdits~\cite{summedits}, are collections of different task-specific benchmarks, developed to address this challenge.
By providing consistent testing harnesses~\cite{AgentBench, bigbench}, composite benchmarks allow practitioners and researchers to understand an LLM's model relevance to open-ended applications.

\vspace{0.05in}
\noindent
\textbf{Conclusion.}
Our analysis of OpenCompass determined that almost all benchmarks are dedicated to testing model correctness and relevance to certain application domains. 
Benchmarks for privacy, security, robustness, efficiency, interpretability, and fairness do exist outside of OpenCompass~\cite{wang2024decodingtrust}, though more work is needed to increase their adoption and coverage.
To answer \textbf{RQ1}, benchmarks alone cover a few topics of the ML testing taxonomy sufficiently.



\vspace{0.05in}
\noindent
\textbf{Threats to Validity.}
We have considered only one benchmark collection. 
It is by far the most popular on GitHub, and therefore representative of today's state of practice for benchmarks dedicated to LLM testing. 
OpenCompass focuses on general LLM capabilities, and as a result benchmarks focused on specific use-cases, such as CodeXGLUE \cite{lu2021codexglue} are not considered.

\section{Open-Source Testing Tools}
\label{sec:tools}

\begin{table*}[t]
\caption{
Open-source tools covering the different topics of ML testing.
Tools analysed in March 2024, with coverage based on the tool's official documentation.
}
\label{tab:coverage_matrix}
\scalebox{0.85}{
\begin{tabular}{@{}lcllllllllllllllll@{}}
                \\
                \\
                \\
                Tool & GitHub Stars& \myrotcell{Debug and repair} & \myrotcell{Test prior. and reduct.} & \myrotcell{Test adequacy generation} & \myrotcell{Test oracle identification} & \myrotcell{Test input generation} & \myrotcell{Framework testing} & \myrotcell{Learning program testing} & \myrotcell{Data testing} & \myrotcell{Privacy} & \myrotcell{Interpretability} & \myrotcell{Fairness} & \myrotcell{Efficiency} & \myrotcell{Robustness and security} & \myrotcell{Model relevance} & \myrotcell{Correctness} & \\ 
             
\cline{1-18}
LangChain Evals~\cite{langchain}& 79674 &  \emptycirc        & \emptycirc                         & \emptycirc                & \fullcirc               & \emptycirc             & \emptycirc         & \emptycirc                & \emptycirc    & \emptycirc & \halfcirc         & \halfcirc  & \emptycirc  & \emptycirc               & \emptycirc       & \fullcirc  & \textbf{Legend}:   \\
Giskard~\cite{giskard}& 2705&  \emptycirc        & \emptycirc                         & \emptycirc                & \halfcirc               & \fullcirc              & \emptycirc         & \emptycirc                & \emptycirc    & \emptycirc & \emptycirc        & \emptycirc & \emptycirc  & \fullcirc                & \emptycirc       & \fullcirc  & \fullcirc: focus of the tool, 3+ tests or features  \\
promptfoo~\cite{promptfoo}& 2141& \emptycirc        & \emptycirc                          & \emptycirc                & \fullcirc               & \fullcirc              & \emptycirc         & \emptycirc                & \emptycirc    & \emptycirc & \emptycirc        & \emptycirc & \halfcirc   & \emptycirc               & \emptycirc       & \fullcirc &\halfcirc: addressed by 1-2 tests or features  \\ 
TruLens-Eval~\cite{trulens}& 1456& \emptycirc        & \emptycirc                         & \emptycirc                & \fullcirc               & \emptycirc             & \emptycirc         & \emptycirc                & \emptycirc    & \halfcirc  & \emptycirc        & \halfcirc  & \emptycirc  & \emptycirc               & \emptycirc       & \fullcirc    &\emptycirc: no related functionality in the tool \\
DeepEval~\cite{deepeval}& 1416 & \emptycirc        & \emptycirc                         & \emptycirc                & \halfcirc               & \emptycirc             & \emptycirc         & \emptycirc                & \emptycirc    & \emptycirc & \emptycirc        & \halfcirc  & \halfcirc   & \emptycirc               & \emptycirc       & \fullcirc    \\ \cline{1-18}


\end{tabular}}
\vspace{-0.1in}
\end{table*}
Open-source tools give practitioners access to testing techniques established in research. In this section, we align the functionalities of open-source tools with topics outlined in our taxonomy (Fig.~\ref{fig:taxonomy}) to pinpoint gaps in applying LLM testing research in practice. 


\vspace{0.05in}
\noindent
\textbf{Tool Selection.} 
Aiming to identify popular open-source testing tools used by practitioners,
we searched for ``llm evaluation" on Github and then ranked the results by the number of GitHub stars, which indicates practitioner interest and usage. 
We only considered repositories with a focus on testing and evaluation, excluding model zoos, tools for running benchmarks, and other LLM repositories without prominent testing packages.
We also limited our analysis to generic tools for text-in, text-out applications that are  applicable to a  variety of LLM deployments, such as summarization, question-answering, and text generation.
The search yielded eight LLM testing repositories with over 1000 stars and proper documentation, of which three contained only functionalities implemented by other repositories and thus were excluded.    Tbl.~\ref{tab:coverage_matrix} lists the selected five tools  and the taxonomy topics (see Fig.~\ref{fig:taxonomy}) they address.

\vspace{0.05in}
\noindent
\textbf{Strengths of Open-Source Tools.} \emph{Test oracle identification} and \emph{tests for correctness} were addressed by all tools studied, highlighted with green in Fig.~\ref{fig:taxonomy}.
\emph{Test oracles} have been a common focus of many open-source tools, including regular expressions or other naive string matching functions~\cite{promptfoo, langchain}. 
Additionally, LLMs or Retrieval Augmented Generation (RAG) systems have been used as test oracles for testing LLMs, i.e., ``LLM-as-a-judge" or ``GEval"~\cite{geval}. 
The focus on test oracles arises from the fact that the input and output of LLMs consist of unrestricted natural language text. This transforms the task of identifying the correct output into a natural language understanding challenge, involving interpreting whether the output conveys the intended meaning~\cite{geval}.  In terms of \emph{correctness},
%
%
%
%
LLM outputs can be considered incorrect for multiple nuanced reasons, based on the requirements of the system. 
For example, since both politeness and factual accuracy are important for the correctness of a customer service LLM, responses that are rude but valid, or polite but wrong are considered incorrect. 
Tests have been implemented to check that responses contain specific, factual information and no hallucinations~\cite{openaievals, promptfoo, trulens, deepeval}. 
Additionally, there are tests dedicated to checking abstract writing properties, such as "concise", "uncontroversial", or "sensitive"~\cite{langchain}. Finally, there are tests for checking whether the output conforms to a specified structure and format~\cite{langchain, promptfoo}, for example, JSON.

\vspace{0.05in}
\noindent
\textbf{Gaps in Open-Source Tools.} Several research topics were not addressed by any of the studied tools.  We denote them by empty circles in Tbl.~\ref{tab:coverage_matrix}. Out of them, model relevance has been addressed by benchmarks (see Sec.~\ref{sec:benchmarks}). 
%
None of the surveyed tools have implemented infrastructure to prioritize tests or skip redundant tests,  
despite the inference of LLMs with millions of parameters being very expensive. 
Therefore, test reduction and prioritization for LLMs remains a promising area for future work, both in research and development.
%
%
Additionally, the original training corpora of the LLMs are often unknown to LLM practitioners, and presumably immense in size, posing challenges for data, learning program, and framework testing.
%
%
%
%
Nevertheless, LLM practitioners and researchers can still debug LLMs with tests and repair bugs with fine-tuning, either through an API~\cite{openaifinetune} or locally with custom-built corpora and learning programs~\cite{runaifinetune, peft}.
However, these workflows are not implemented by any tools we studied, and thus are important directions for future work. 


\vspace{0.05in}
\noindent
\textbf{Conclusion.}
By mapping open-source tools to topics on the taxonomy, we discovered that test oracle identification and correctness are well-addressed by open-source tools, even though comparisons between different test oracles for different tasks are yet to be made.  Furthermore, most other ML testing topics remain unaddressed, namely, debugging and repair, test prioritization and reduction, test adequacy generation, framework testing, learning program testing, data testing, and model relevance. 
To answer \textbf{RQ1} and \textbf{RQ2}, open-source tools cover only a small portion of the topics in the ML testing taxonomy, and this lack of tooling directly prevents practitioners from effectively applying LLM testing research.

\vspace{0.05in}
\noindent
\textbf{Threats to Validity.}
Our tool survey consists of only five of today's most popular tools, selected though only one search query on GitHub, and thus is clearly incomplete.
Furthermore, the functionalities of the tools are determined solely based on documentation, potentially providing limited insights compared to thorough experimentation and code review.

\section{Practice of LLM Testing}
\label{sec:developers}

In this section, we examine how LLM testing is done in practice by analyzing testing-related discussions on online forums. By comparing discussions with topics from our taxonomy and testing tools, we identify potential gaps in current testing practices. 

\vspace{0.05in}
\noindent
\textbf{Data.} We focused on Reddit, which is one of the most popular online forums \cite{eghtesadi2020facebook}, because its subreddits are issue-oriented forums widely used and accessible to a broad audience \cite{krohn2022subreddit}. Specifically, we selected the subreddit r/LocalLlama because of its singular focus on LLMs and popularity on the website -- since its creation in March 2023, it has  become one of the most popular subreddits on the forum, ranking among the top five percent in size\footnote{Although the stated objective of this subreddit is to "discuss about Llama, the large language model created by Meta AI", in practice, posts range from topics such as technology policy, news, model and code development, to popular LLMs released by firms such as OpenAI and Anthropic.}. Subreddit submissions and their corresponding comments (from March 2023 to January 2024) were collected from Academic Torrents~\cite{torrents1, torrents2}. 
To characterize conversations occurring on the platform, submissions and comments were combined, such that the unit of analysis is a subreddit thread. The final data set consisted of 15,209 submissions and 11,377 comments across the subreddit, which amounted to 11,344 threads\footnote{\# of threads is greater than \# submissions since the threads account for partial conversations where the submissions may have been removed from the data.} (submissions and corresponding comments).

\vspace{0.05in}
\noindent
\textbf{Testing-related Keywords.}
To examine how subreddit users from diverse backgrounds discuss LLM testing, we first searched the data for keywords associated with testing software systems \cite{braiek2019tfcheck, grosse2014testing, murphy2008properties, xie2011testing, dwarakanath2018identifying}, machine learning \cite{goodfellow2014explaining,kurakin2016adversarial,madry2017towards,tramer2017ensemble}, LLMs \cite{wang2024software}, and the names of the testing tools  in Tbl.~\ref{tab:coverage_matrix}. These keywords were derived from the corresponding literature, and thus differ from the taxonomy proposed in Fig.~\ref{fig:taxonomy}. Tbl.~\ref{tab:table_3} reports the frequency and percent of the testing-specific keywords found in the data. 
Keywords related to testing appeared in approximately 0.8\% of the total threads.
Next, ``unit test" (N=38) and notions of ``toxicity” (N=29) are the most frequently discussed testing keywords, 
while other traditional software (e.g., ``test oracle" (N = 0), ``functional test” (N = 0)) and machine learning (e.g.,``adversarial attack” (N = 5), ``adversarial example” (N = 0)) testing keywords appeared in less than 0.05\% of the data, if  at all. The testing tools  in Tbl.~\ref{tab:coverage_matrix} similarly proved unpopular amongst subreddit users and collectively appeared in less than 0.02\% of the data.

\vspace{0.05in}
\noindent
\textbf{Analysis of Discussions.} To better characterize the testing-related conversations, we conducted semantic thematic analysis~\cite{clarke2021thematic} for identifying and analyzing patterns, themes, and insights within data containing testing-related keywords. 
The quality of these conversations varied across posts and ranged from general discussion of testing, to particular methods for doing so in practice. 
We noticed that the former was particularly obvious for ``toxicity", which was frequently mentioned in discussions of toxic or biased model outputs and performance reports of newly released models. 
In contrast, discussion of unit testing proved more actionable with many users reporting the tests they personally developed to evaluate models or those reported in the literature. Despite the relative popularity of unit testing on the subreddit, there appeared to be a lack of coherent guidance or steps on how to practically test LLMs.
For example, in response to a post asking ``Prompt Engineering Seems Like Guesswork - How To Evaluate LLM Application Properly?", some users responded by noting methods such as unit tests, while others reported processes they independently developed. For example, the quote in panel A of Tbl.~\ref{tab:quotes} notes a combinatorial testing approach with user evaluation. In contrast, panel B highlights a hierarchical method developed by another user which seemingly draws from both software engineering  (e.g, unit testing and user acceptance testing)  and ML practices (e.g., evaluation on benchmarks).

\begin{table}[t]
    \centering
    \caption{Example Quotes from r/LocalLama.} \label{tab:quotes}
    \vspace{-0.1in}
   \scalebox{0.72}{
    \begin{tabular}{|p{5cm}|p{6cm}|}
    \hline
    A. Example of Combinatorial Testing & B. Example of a Hierarchical Testing Method\\
    \hline
       \textit{``This is what I did. Write a set of prompts in the domains I’m interested in, a list of models, a list of generation presets. \textbf{Run all combinations (prompts x models x presets) of settings, then rate outputs on a 4-point scale.} I’m doing it the hard way and manually rating outputs. Then I can calculate relative scores for specific domains, models, presets, or any combination of them, and see what works best for what....''}  & \textit{``I've been working in a similar setup (chatbot evaluation) for my company and our take is to \textbf{consider 3 tiers of evaluations}:
       \begin{itemize}
           \item Tier 1: evaluate predicted vs expected across a reference dataset, using classical textcomparison metrics... and semantic embedding similarity...
           \item Tier 2: evaluate prediction against expected properties (ex: schema compliance, lack of toxicity, robustness, factuality)...
           \item Tier 3: Look at user feedback (implicit or explicit) to see how your model is performing vis a vis the real user experience...''
       \end{itemize}
       } \\
         \hline
    \end{tabular}}
    \vspace{-0.15in}
\end{table}

\vspace{0.05in}
\noindent
\textbf{Conclusion.} 
The apparent novelty of many of the methods noted on r/LocalLlama suggests both the utility of SE  and ML approaches to testing LLMs, and potential discrepancies between theory and practice. 
In addition to explicitly drawing on methods such as unit and integration testing to test LLMs, the methods proposed by users are ad-hoc and experimental.
Interestingly, we noticed that the discussions do not explicitly appeal to SE or ML terminologies. Thus, the mentioned testing methods differ from both the academic literature and the testing tools  in Tab.~\ref{tab:coverage_matrix}. Our results show a large gap between research and practice, which highlights the necessity for better communication between researchers and practitioners. 
\vspace{0.05in}
\noindent
\textbf{Threats to Validity.}
The analysis of the subreddit, r/LocalLama, serves as a preliminary examination of user discussions of LLM testing.  Specifically, the search terms used in the analysis were limited, potentially resulting in the omission of relevant keywords pertinent to testing discussions. Additionally, the diverse user group on the subreddit may not fully represent professionals actively engaged in the testing field, thus impacting the generalizability of the findings. 

\begin{table}
\caption{Frequency and Proportion of Keywords in Data. }
    \centering
     \vspace{-0.1in}
    \scalebox{0.67}{
    \begin{tabular}{ccc| ccc}
    \toprule
    \multicolumn{3}{c|}{Testing Approaches} & \multicolumn{3}{c}{Testing Tools}\\
    \midrule
         Keyword\emph{$^{a}$}&  Frequency& Percentage\emph{$^{b}$} &  Keyword&  Frequency& Percentage\emph{$^{a}$}\\
        \midrule
        
        unit test           &           38 &        0.33 & promptfoo           &            1 &        0.01\\ 
        
        toxicity            &           29 &        0.26 & TruLens-Eval        &            1 &        0.01 \\
        integration test    &            8 &        0.07  & LangChain Evals     &            0 &        0.00   \\
        conciseness         &            6 &        0.05  & Giskard             &            0 &        0.00  \\
        adversarial train   &            5 &        0.04 & DeepEval            &            0 &        0.00 \\
        adversarial attack  &            5 &        0.04 \\
         \bottomrule
    \end{tabular}}
    \smallskip \\
    \footnotesize
    \emph{$^a$}Keywords \emph{adversarial example/sample}, \emph{mutation/functional/metamorphic/property based/bias test}, \emph{test oracle} and \emph{neuron coverage} in the Testing Approaches have 0 percentage and thus are omitted.\\
      \emph{$^b$}Percent of the total number of subreddit threads (N = 11,344).\\
    \label{tab:table_3}
    \vspace{-0.2in}
\end{table}

\section{Conclusion and Future Outlook}
\label{sec:future}
In this paper, we proposed a vision for organizing and unifying efforts on LLM testing from both research and industry practices through a taxonomy of ML testing in SE. We presented a preliminary study on current research methods, open-source benchmarks, testing tools, and online discussions by practitioners.

Our results allow us to answer \textbf{RQ1}: a large portion of LLM testing research is not yet incorporated into publicly available tools and benchmarks, leading to future engineering and research directions. Regarding the answer to \textbf{RQ2}, we noticed that developers online do not explicitly reference SE or ML testing topics but are incorporating SE testing techniques into their practices. Although preliminary, these results show promising prospects for organizing existing methods using taxonomies such as the one in Fig.~\ref{fig:taxonomy}, identifying gaps and future directions, and promoting communication and collaboration between researchers and industry practitioners. 




Our study has many limitations:  not only did we analyze only \emph{some} benchmarks, tools and online discussion forums, the pace of change of SE engineering will soon render our March 2024 snapshot and conclusions based on it, obsolete.   Future work needs to increase the scope of research publications, tools, benchmarks and practitioner online discussions (and also interview studies) as well as make frequent updates to the knowledge gathered.
As applications of LLMs continue to grow, in 2030, the taxonomy of LLM testing should continue to expand, helping to connect research and industry practices. Structuring tools and benchmarks following this taxonomy facilitates user access to testing techniques tailored to their application requirements. Ultimately, we hope our work will facilitate testing of LLMs in a clear, principled, and comprehensive way, enabling usage of LLMs in high-stake or safety-critical systems.

\section*{Acknowledgement}
The authors would like to thank the reviewers of Software Engineering 2030 for their valuable feedback and comments.

\bibliographystyle{ACM-Reference-Format}
\bibliography{ref}

\end{document}